# Retracted papers by Iranian authors:

# Causes, journals, time lags, affiliations, collaborations

Ali Ghorbi[1*], Mohsen Fazeli-Varzaneh[1], Erfan Ghaderi-Azad[1],

Marcel Ausloos[2, 3, 4], Marcin Kozak[5]

Ali Ghorbi ORCID: 0000-0003-1411-575X

Mohsen Fazeli-Varzaneh ORCID: 0000-0001-8614-8513

Erfan Ghaderi-Azad ORCID: 0000-0002-8469-0234

Marcel Ausloos ORCID: 0000-0001-9973-0019

Marcin Kozak ORCID: 0000-0001-9653-3108

1. Department of Knowledge and Information Science, Faculty of Management, University of Tehran, Tehran, Iran
2. School of Business, University of Leicester, Brookfield, Leicester, LE2 1RQ, United Kingdom
3. GRAPES, rue de la Belle Jardinière, 483/0021, B-4031, Liège Angleur, Belgium.
4. Department of Statistics and Econometrics, Bucharest University of Economic Studies, Calea Dorobantilor 15-17, Bucharest, 010552 Sector 1, Romania
5. Department of Journalism, Media and Social Communication, University of Information Technology and Management in Rzeszów, Sucharskiego 2, 35-225 Rzeszów, Poland

*Corresponding author email: alighorbi73@ut.ac.ir

## Abstract

This study aims to analyze 343 retraction notices indexed in the Scopus database, published in 2001–2019, related to scientific articles (co-)written by at least one author affiliated with an Iranian institution. In order to determine reasons for retractions, we merged this database with the database from Retraction Watch. The data were analyzed using Excel 2016 and IBM-SPSS version 24.0, and visualized using VOSviewer software. Most of the retractions were due to fake peer review (95 retractions) and plagiarism (90). The average time between a publication and its retraction was 591 days. The maximum time-lag (about 3,000 days) occurred for papers retracted due to duplicate publications; the minimum time-lag (fewer than 100 days) was for papers retracted due to "unspecified cause" (most of these were conference papers).

As many as 48 (14%) of the retracted papers were published in two medical journals: *Tumor Biology* (25 papers) and *Diagnostic Pathology* (23 papers). From the institutional point of view, Islamic Azad University was the inglorious leader, contributing to over one-half (53.1%) of retracted papers. Among the 343 retraction notices, 64 papers pertained to international collaborations with researchers from mainly Asian and European countries; Malaysia having the most retractions (22 papers).

Since most retractions were due to fake peer review and plagiarism, the peer review system appears to be a weak point of the submission/publication process; if improved, the number of retractions would likely drop because of increased editorial control.

**Introduction**

In the 21st century, academia has been faced with various and numerous instances of misconduct. Many authors try to exploit the academic publishing system in order to publish articles that would not otherwise be accepted for publication. In that way, they aim to achieve unearned success for many possible reasons, such as career advancement, obtaining additional gratification, gaining credibility or respect in scientific society, or simply boosting their self-image. These days, with millions of authors trying to publish their research and many of them succeeding, scientists face increased expectations. What twenty years ago constituted a great achievement in terms of the number of published articles can now be considered barely the minimum, or even below the minimum "for scientific success". Thus, publishing ten instead of five papers a year is a tempting prospect; some scientists eventually surrender to social or academic pressure and choose an unethical path.

The scientific community had to react to this phenomenon, and it did, both as a group of individuals and as a community. As examples of individual reactions, we include reviewers, most of whom now pay much attention to any unethical behavior of the authors of the manuscripts they review, as well as journal editors and technical staff who guard the quality of scientific journals. Many scientists also discourage and stigmatize unethical behavior of their peers.

As examples of community reactions, we can mention conferences, journals and organization bodies dedicated to ethics in science. Among such bodies, the Committee on Publication Ethics (COPE) plays a crucial role. A nonprofit organization consisting of editors and publishers who work to improve the quality of scholarly publishing, COPE creates standards and definitions, as well as educates and supports editors, publishers and all those involved in publication ethics.

The review processes as currently organized by journals do not fully control the ethical quality of publications (Bornmann and Mungra, 2011). One of the tools used to overcome this issue—albeit a *post mortem* tool—is article retraction. A published article that later proves to be unworthy of publication for particularly serious reasons may be retracted. Surely, since this is a serious tool, it must be used with caution so that innocent people are not affected. COPE suggests that journal editors should consider whether or not to retract an article in one of the following four situations: (i) the editors have an evidence that the findings are unreliable, either as a result of misconduct or honest error; (ii) the same findings have already been published and this fact was not properly addressed (e.g., by obtaining permission or providing a reference); (iii) the article plagiarises another published document; or (iv) the article is based upon unethical research (Wager et al., 2009). Based on such COPE guidelines, an article could be retracted due to "honest errors," "misrepresentation", "data manipulation," "poor data management" (such as the author being unable to generate data to support his or her results), "plagiarism" (including self-plagiarism), "duplication of text," or "failure to disclose conflicts of interest" (Wager et al., 2009).

It seems that the retraction phenomenon is country-dependent, since some national scientific environments put more emphasis on quality and others on quantity. When quantity of scientific output becomes more important than its quality, it seems much easier (and likely more tempting) to exploit this situation through some form of unethical behavior (like plagiarism). Recently, Campos-Varela and Ruano-Raviña (2019) analyzed 2013–2016 PubMed-indexed publications in terms of retractions; from their results, it follows that Iran was the inglorious leader of what we could call the retraction business. Dakhesh and Hamidi (2020) mentioned various measures that were (recently) taken in Iran to overcome this issue, and suggested that "It's obvious that these authors and the two mentioned journals cannot be a true representative for Iranian researchers".

Yet, given this escalation of the problem in Iran, this present paper reports studies on the retraction problem among Iranian-affiliated authors. We do so by examining various aspects of retracted papers, mainly focusing on who wrote them, where they were published, and why they were retracted. The knowledge gained from such analyses will hopefully contribute to (i) the general knowledge of the retraction problem; and (ii) specific knowledge on Iranian retractions, a particularly important issue given previous results suggesting that Iran might be the greatest contributor to retractions, something in turn suggesting that Iranian science may be suffering from a serious issue of unethical behavior.

**Literature review**

To set up the frame of the present research and to answer specific research questions, we outline here the relevant, mostly recent, papers. The literature on retraction is already vast, and one of the first documents triggering increasing interest in the phenomenon is Stewart and Feder's (1987) comment on a problematic paper. The need to study retractions and publish the results of such research quickly appeared a key for the process of dissemination of scientific results.

A literature survey indicates that retractions constitute a growing problem for science. Wager and Williams (2011) showed that the tendency for retracting articles increased rapidly in the 1980s. Various pictures of the retraction phenomenon follow from various studies, depending on the time window, publication source, but also other aspects (Lei and Zhang, 2018). Tang et al. (2020) found that "China stands out with the fastest retracting speed compared to other countries." From Campos-Varela and Ruano-Raviña's (2019) study, however, it follows that "the highest proportion of retracted publications corresponded to Iran (15.52 per 10,000), followed by Egypt and China (11.75 and 8.26 per 10,000)."

Steen (2011) examined 788 retracted articles from 2000 to 2010. He observed that the top five countries in terms of the number of retracted articles were the United States (260), China (89), Japan (60), India (50), and England (45). Note that Iran was *not* among those countries; this is a different result from that observed in Campos-Varela and Ruano-Raviña (2019) conclusion about Iran's contribution to retracted papers. Callaway (2016) stressed that 58 papers by Iranian researchers were retracted mainly due to authorship manipulation, peer review manipulation, and plagiarism. Later, Dakhesh and Hamidi (2020), in their comment on Campos-Varela and Ruano-Raviña's (2019) paper, stressed that Iran took serious actions against its authors whose papers had been retracted. This led the authors conclude that retractions of Iranian papers, although a serious problem in the past, should not constitute a so serious problem anymore.

Wager and Williams (2011) determined that approximately 0.02% of all publications listed in Medline from 2005 to 2009 had been retracted. Based on 312 retracted articles from 1988 to 2008, Wager and Williams (2011) concluded that 63% of the retractions were due to the requests from the authors of these retracted articles. From Vuong's (2020) study, a different picture follows: 53% of the retraction notices (for 2046 retracted papers published between 1975 and 2019) are not specifying who initiated the retraction.

Another important aspect of scientometrics research on retractions pertains to reasons behind retractions. Generally, the causes are divided into three categories: research misconduct, scientific errors, and moral and/or political reasons (Aspura et al., 2018; Wager and Williams, 2011). For example, Budd et al. (1998) examined 235 retracted articles from 1966 to 1997 indexed in Medline. In their study, retraction reasons included most of all scientific errors (91 papers, 38.7%) and research misconduct (86 papers, 36.5%), but much less frequently unrepeatable results (38 papers, 16.1%) as well as no clear reason (20 papers, 8.5%). Similar results were reported by Wager and Williams (2011), Fang et al. (2012), Grieneisen and Zhang (2012), Ribeiro and Vasconcelos (2018), Dal-Ré and Ayuso (2019), and Elango et al. (2019). In Chauvin et al.'s (2019) study, the most common reasons for retractions in the field of emergency medicine were plagiarism (29%) and duplicate publication (11%); similar observations were made by Elango et al. (2019) for Indian retractions.

Elango et al. (2019) analyzed the types of retracted papers. Out of 239 India-affiliated retracted articles indexed in Scopus between 2005 and 2018, 82% were published in journals; the remaining ones were in conference proceedings. Ghorbi (2019) analyzed citations and thematic classification of retracted papers affiliated to the Middle East countries, and showed that about half of Iranian retracted papers were categorized in Life Sciences & Biomedicine, among which most papers were in the Oncology and Pathology sub-categories.

One of the most important aspects related to any unethical behavior pertains to country-wise differences and international collaborations. One might presume that most retractions should come from countries whose science is at a low level, but research shows that this is not true: scientifically strong countries, with diverse and abundant scientific output, tend to have more retracted publications (Zhang et al., 2019). Ghorbi and Fahimifar (2020) studied the status and collaboration patterns of all retracted papers indexed in Web of Science during 1997–2018; they compared institutions and countries based on the absolute number of retracted papers. With 205 retracted papers, Iran took the 7th place in terms of the number of retracted papers. However, the Islamic Azad University of Iran was the second institution with the most retracted papers in the world. From Ghorbi and Fahimifar (2020) study, it follows that indeed, at least until 2018, Iran was among the countries with a serious retraction problem.

**Research objectives**

The main question behind this research—which we built based upon the literature and reasoning presented in the Introduction—is that Iranian science suffers from a serious retraction problem (as suggested in various sources), but also that the situation has improved recently (as suggested—or rather hoped—by Dakhesh and Hamidi (2020)).

We wish to emphasize that despite dealing with a sensitive topic related to unethical behavior of scientists, this research is fully objective and is not politically motivated, partially because the paper is the result of an international collaboration. The research and the paper aim to contribute to the global bibliometric and scientometric knowledge by providing partial knowledge, related to one particular country. This, we hope, will help to improve the behavior of the global scholarly publishing community. If this work addresses one country—The Islamic Republic of Iran—, it is because frequent results have pointed out that this country has strongly contributed to the global retraction problem. This particular observation calls for a detailed analysis of the retraction phenomenon in Iranian science, something that this paper offers.

**Materials and methods**

On 20th January 2020, we queried the Scopus database to identify retracted publications using the search term "retract*", with an additional country filter, to choose Iran-affiliated retraction notices that were published from 2001 to 2019. This search strategy, previously used by Elango et al. (2019) and Aspura et al. (2018) for other countries, yielded a total of 418 retractions. To filter out non-retractions (that is, scientific articles in whose titles a term "retract*" appeared that were not related to retraction), we checked all these papers, a process that led us to detect 343 actual retractions. For further analyses, we extracted the data as a CSV file.

Besides standard summary statistics, we also applied the Modified Collaboration Coefficient (MCC) (Savanur and Srikanth, 2010), based on the Collaboration Coefficient (de Solla Price and Beaver, 1966; Ajiferuke et al., 1988), calculated for a particular set of articles as follows:

$$MCC = \frac{A}{A-1}\left\{1 - \left[\sum_{j=1}^{k}\left(\frac{1}{j}\right) \times \frac{f_j}{n}\right]\right\} \qquad (1)$$

where $f_j$ is the number of papers with $j$ co-authors, $n$ is the total number of papers published in the investigated set (e.g., a category or a country), $k$ is the largest number of authors per paper in the set, and $A$ is the total number of authors in the set.

MCC ranges between 0 and 1; an MCC value close to zero represents a trend toward single-author articles; a value close to 1 represents a trend toward multi-author articles.

We analyzed Iranian retractions in the following contexts:

- time lag (Fang et al., 2012);
- journals (Steen, Casadevall and Fang, 2013; Bornemann-Cimenti, Szilagyi, and Sandner-Kiesling, 2016; Steen, 2012) and their quality measures, like impact factor (Fang et al., 2012);
- the size/level of institution (Stavale et al., 2019); and

- retraction reasons (e.g., Budd et al., 1998; Wager and Williams, 2011; Fang et al., 2012; Grieneisen and Zhang, 2012; Ribeiro and Vasconcelos, 2018; Dal-Ré and Ayuso, 2019; Elango et al., 2019; Chauvin et al., 2019), including retractions requested by the authors themselves (Wager and Williams, 2011).

To study retraction reasons and determine the share of Iranian retractions in global retractions, we also extracted data from another data source: the Retraction Watch database. The advantage of this database for this type of analysis is that it already includes reasons behind retractions. Therefore, all analyses—including ours—of retractions in terms of retraction reasons based on Retraction Watch are comparable. The Retraction Watch database contains more (728) Iranian retractions than we found in Scopus (343). Therefore, for a general analysis of the share of Iranian retractions in global retractions, we used the whole database from Retraction Watch. However, to analyze retraction reasons of the 343 retractions we found in Scopus, we merged these two databases, keeping only those retractions that were included in both databases. Fortunately, all the retractions we retrieved from Scopus were included in the Retraction Watch database; in so doing we added reasons to all the 343 retractions found from Scopus.

We independently reviewed these reasons based on the COPE guidelines, and then categorized the reasons. If any inconsistencies occurred, we resolved them by discussion and final agreement. This way, we obtained the following eleven categories of retraction reasons: fake peer review, plagiarism, duplicate publication, authorship problems, scientific error, publisher error, falsification/fabrication, copyright claims, conflict of interest, unspecified, and other.

Data analysis was performed using Excel 2016 and IBM-SPSS version 24.0. To visualize the international collaboration map, we used VOSviewer.

**Results**

Table 1 summarizes the data from the Retraction Watch database, by comparing those for Iran with those for the whole world; Figure 1 shows the share of Iranian retractions in global retractions over time (also based on Retraction Watch data). In 2016, Iran encountered quite a problem with retractions, with as high a share as 14% of all retracted papers in the world having been affiliated with at least one author from an Iranian institution. Given such a small country in terms of scientific output, this is a huge number. Nonetheless, we notice that this share dropped to 6% in 2019.

Table 1. Main statistical characteristics of the Retraction Watch database in terms of retracted papers across 2001–2019 in the world and those having at least one affiliation with the Islamic Republic of Iran (IRI). Data source: Retraction Watch.

| No. of retractions | Min | Max | Total | Mean | Median | SD | Skewness | Kurtosis |
|---|---|---|---|---|---|---|---|---|
| **World** | 33 | 4,918 | 18,840 | 991.6 | 796 | 1,197 | 2.14 | 4.38 |
| **IRI** | 0 | 120 | 728 | 38.3 | 29 | 41.8 | 0.74 | -0.88 |

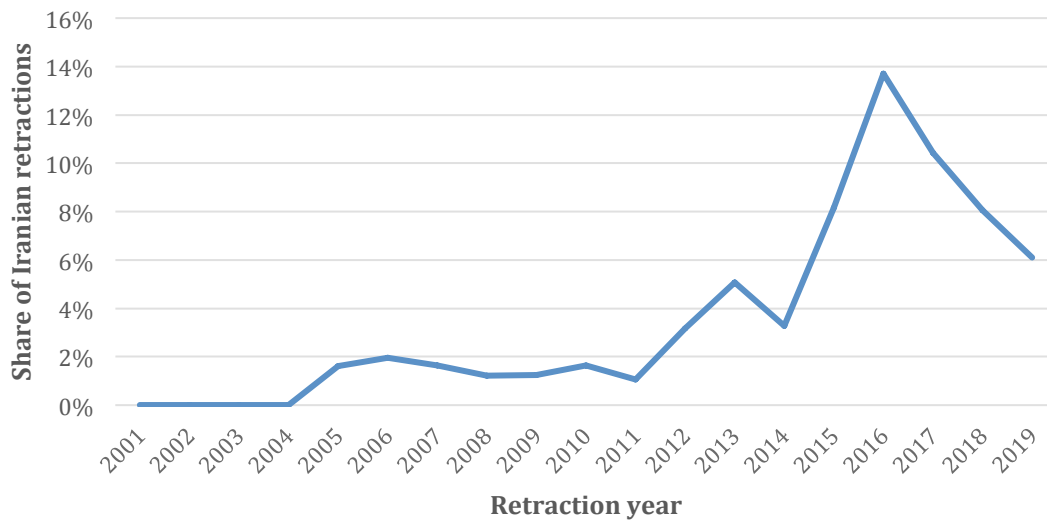

Fig 1. The share of retracted Iranian documents in the global number of retracted documents (2001–2019). Data source: Retraction Watch; the graph is prepared based on all 728 Iranian retractions found in Retraction Watch.

Iranian science shows a well-known global trend of rapidly increasing publication output, with more and more Iranian papers being published each year (Figure 2). A naïve simple linear regression analysis well describes this trend, with the determination coefficient being equal to 98.1%. Although we cannot extrapolate this trend into the future, it does show that in the twenty-first century, Iranian science has been facing a rapid development, just like world science.

On one hand, the number of retracted papers by Iranian-affiliated authors has been showing an increasing trend (Figure 3); on the other hand, this trend had turn overs in 2010 and 2016, when the number of retracted Iranian documents was incomparably high. During 2001–2019, 343 Iranian papers were retracted; almost half of them have been retracted in 2010 and 2016. Many papers that were retracted in 2010 were published during that same year, and some of those retracted in the next years (more precisely, 66 of the 75 papers published in 2010 were retracted in that year). The second peak in retracted documents (61) occurred in 2015, one year after a peak in the number of later-retracted publications (70). This shows that there is some "not short" time lag between publication and retraction. On average, papers were retracted 591 days after publication. Conference papers had the shortest time from publication to retraction. For example, in 2015, 64 of the 65 conference papers were retracted in the same year of publication, a result explaining the 2015 peak in Figure 3. The yearly number of documents and the number of retracted papers by Iranian authors showed a significant correlation (the *p*-value of 0.004)[1], but the Kendall correlation coefficient was rather low ($\tau = 0.487$)—so the significance being also due to the size of the sample (Kozak, 2008).

Interestingly, among the 68 papers retracted in 2010, only three—all published in journals—were retracted for plagiarism, while the other 65 were conference papers with no specified retraction reason. As many as 45 of these 65 (69%) retracted conference papers were published by authors affiliated to the Islamic Azad University[2], an institution which had the greatest contribution to Iranian retractions; this institution was pointed out by Ghorbi and Fahimifar (2020) as being the second greatest contributor to global retractions.

The retraction peak in 2016 has a different story. Most of these retractions—48 out of 70—were of papers published in 2013–2016 in two medical journals: *Tumor Biology* and *Diagnostic Pathology*. Note that in Ghorbi's (2019) research, most retracted Iranian papers were from Life Sciences & Biomedicine, especially in the fields of Oncology and Pathology. As Fig.A1 in the Appendix shows, these two journals have faced a high number of submissions from Iran and published many more papers from Iran than they used to do before that time. Nonetheless, after the great retraction action in 2016, both journals almost stopped publishing Iranian research. Worth noticing is that both journals are reputable in their fields, thereby suggesting that even a reputable journal can be a victim of unethical behavior on a large scale. Also, note that both *Tumor Biology* and *Diagnostic Pathology* are open-access journals; this may have induced authors of these articles to make results more quickly visible, but whether open-access journals attract unethical behavior more than other journals needs further additional research, on a wider scale than the present study.

---

[1] As also noticed by Lei and Zhang (2018), for China, for which $\tau = 0.84$.
[2] Note that "Islamic Azad University" has many branches in different cities, like the Islamic Azad University of Isfahan, the Islamic Azad University of Mashhad, the Islamic Azad University of Tabriz, etc. Scopus, however, groups all of them into one institution, the Islamic Azad University, which makes it a large university.

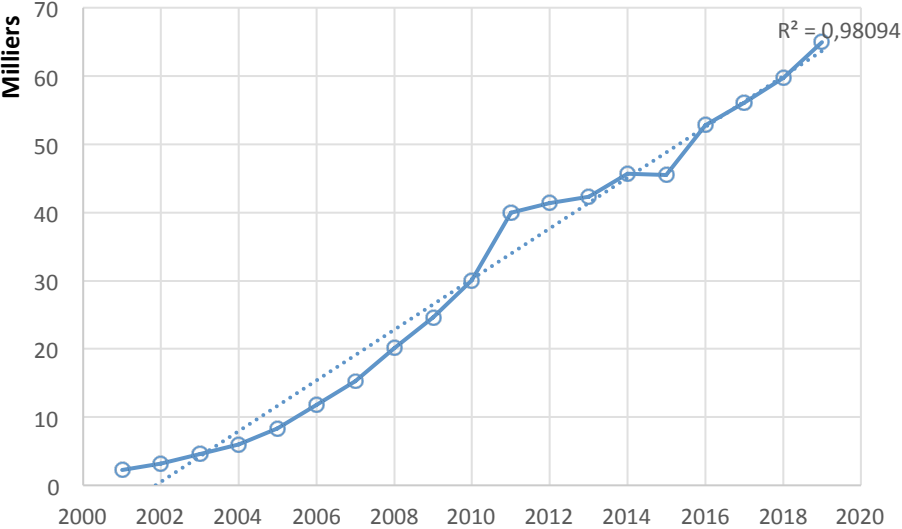

Fig. 2. The total number of documents published by Iranian authors in 2001–2019, with a linear regression line showing a clear trend across this period (based on the average of over 30 thousand papers per year). Data source: Scopus.

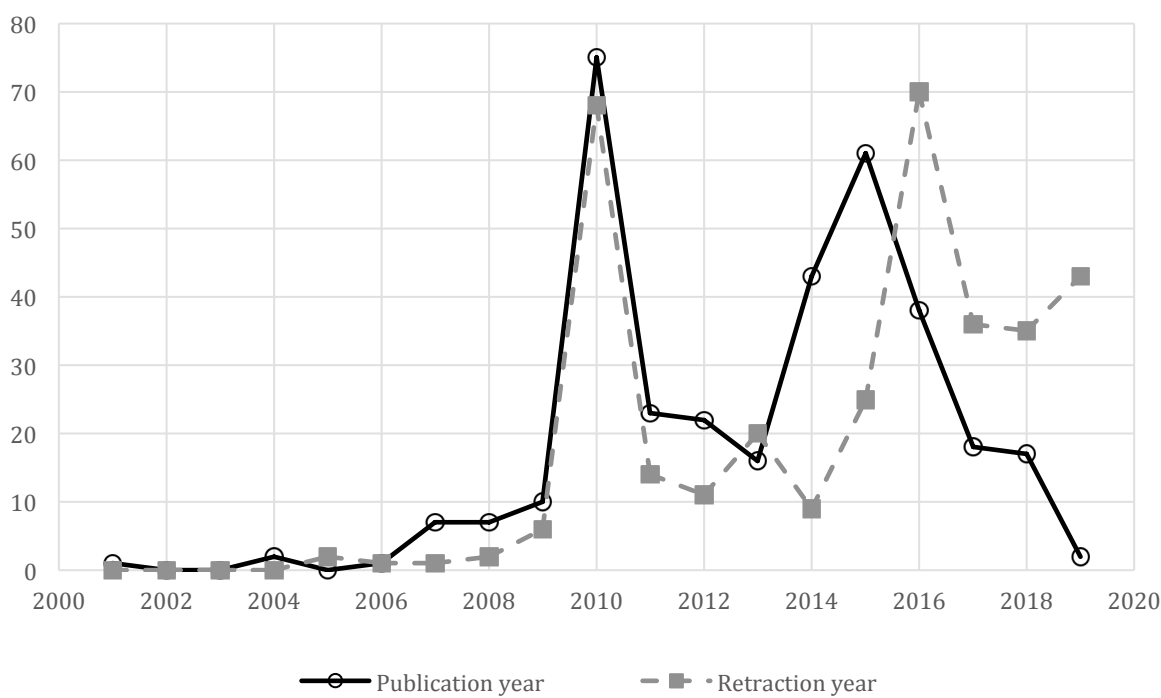

Fig 3. Retracted Iranian papers during 2001–2019: The dashed line shows the number of papers retracted a given year, and the solid lines shows the number of papers published in a given year (limited to retracted papers). Data source: Scopus.

Table 2. Source type and document type of the 343 Iranian retractions. Data source: Scopus.

| Source | Document type[#] | Count | Share |
|---|---|---|---|
| | Erratum | 219 | 63.8 |
| | Article | 28 | 8.2 |
| **Journals** | Retracted | 13 | 3.8 |
| **(264 papers in 138 journals)** | Review | 2 | 0.6 |
| | Editorial | 1 | 0.3 |
| | Note | 1 | 0.3 |
| **Conference Proceedings** **(79 papers in 24 proceedings)** | Paper | 79 | 23.0 |

[#] "Document type" here represents the type of document of a retraction notice, as indexed by Scopus.

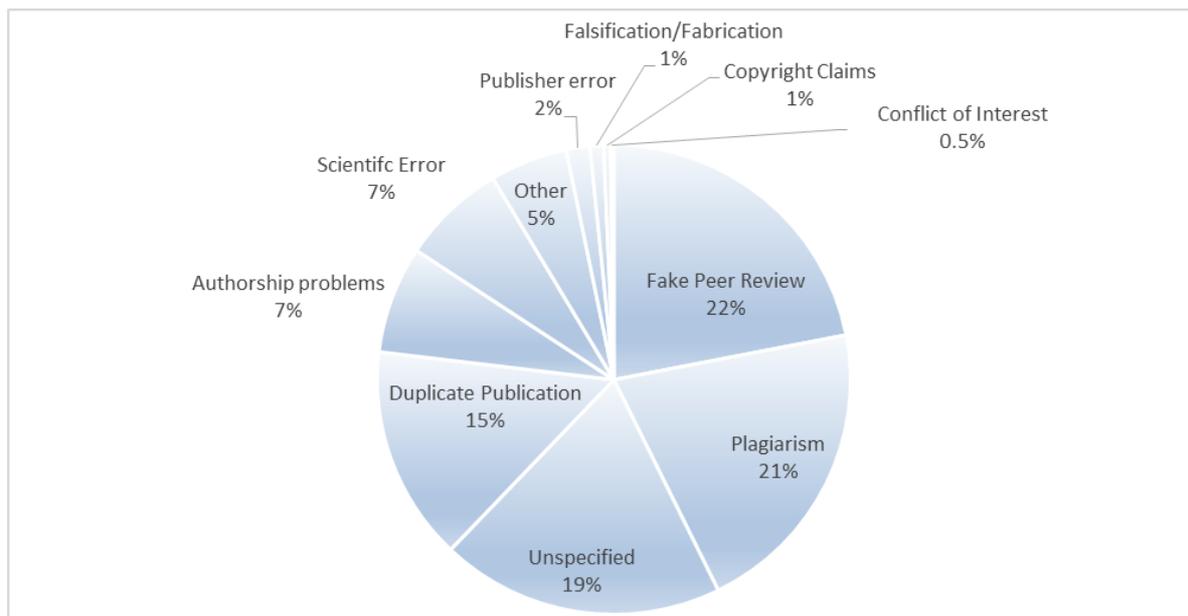

Fig 4. The share of reasons of Iranian retractions. Data source: Retraction Watch; the data include the 343 retractions from Scopus, for which retraction reasons were retrieved from Retraction Watch.

Most retracted papers (264) were published in journals—as many as 138 journals (Table 2). For 219 papers, Scopus classified retractions notices as errata, although "this word was not used in the retraction notices themselves," as already noticed by Elango et al. (2019).

Over half of the journal articles were retracted because of a double cause: "fake peer review" and "plagiarism" (Figure 4). Few were published in conference proceedings, and none of these retractions provided a reason.

Table 3 summarizes retraction reasons (retrieved from Retraction Watch) for the 343 documents. Since some papers were retracted for more than one reason, the reasons sum up to 433; thus the sum of their share is over 100%. The most common retraction reasons were fake peer review (95) and plagiarism (90). Eighty-four retractions did not specify the reasons.

Table 3 also reports the modified collaboration coefficient (MCC). The whole collection had MCC=0.61. Note that the papers retracted due to authorship problems and fake peer review(s) had higher MCC values than those retracted due to plagiarism and duplicate publication. These results should not come as a surprise, since plagiarism and self-plagiarism are less likely to be conducted by large groups of co-authors. Note that out of the 343 retracted Iranian papers, only 36 were written by a single author and 96 by two co-authors.

Table 3. Main reasons for retractions. <P-R> is the average time (in days) between publication date and retraction date; MCC is the Modified Collaboration Coefficient. Share is calculated as the ratio to the total number of retracted papers (343). Data source: Retraction Watch; the data include the 343 retractions from Scopus, for which retraction reasons were retrieved from Retraction Watch.

| Reason for retraction | Count | <P-R> | MCC | Share |
|---|---|---|---|---|
| Fake peer review | 95 | 669 | 0.72 | 27.7 |
| Plagiarism | 90 | 672 | 0.62 | 26.2 |
| Unspecified | 84 | 129 | 0.53 | 24.5 |
| Duplicate publication | 64 | 1057 | 0.58 | 18.7 |
| Authorship problems | 32 | 813 | 0.72 | 9.3 |
| Scientific error | 31 | 745 | 0.70 | 9.0 |
| Other | 23 | 262 | 0.50 | 6.7 |
| Publisher error | 7 | 754 | –[#] | 2.0 |
| Falsification/fabrication | 4 | 476 | – | 1.2 |
| Copyright claims | 2 | 932 | – | 0.6 |
| Conflict of interest | 1 | 276 | – | 0.3 |

[#] We did not determine MCC for sets with fewer than 10 papers, since MCC has limited interpretation in such instances.

Table 4. Top 10 journals with most retractions by Iran-affiliated authors. Data source: Scopus.

| Journals | Count | Share | <P-R> (Days) | CiteScore | SJR | SNIP | Subject Area |
|---|---|---|---|---|---|---|---|
| *Tumor Biology* | 25 | 7.29 | 377 | 3.21 | 1.057 | 0.818 | Biochemistry, Genetics and Molecular Biology: Cancer Research |
| *Diagnostic Pathology* | 23 | 6.71 | 747 | 1.98 | 0.804 | 0.864 | Medicine: Pathology and Forensic Medicine; Medicine: Histology |
| *Progress in Organic Coatings* | 12 | 3.50 | 87 | 3.7 | 0.822 | 1.348 | Chemical Engineering; Chemistry; Materials Science |
| *International Journal of Hydrogen Energy* | 10 | 2.92 | 285 | 4.16 | 1.1 | 1.128 | Energy; Physics and Astronomy |
| *Neural Computing and Applications* | 8 | 2.33 | 443 | 4.2 | 0.637 | 1.481 | Computer Science |
| *Energy and Buildings* | 5 | 1.46 | 2990 | 5.36 | 1.934 | 1.826 | Engineering |
| *Materials Science and Engineering A* | 5 | 1.46 | 663 | 4.62 | 1.778 | 2.015 | Materials Science; Engineering; Physics and Astronomy |
| *Biochemical Genetics* | 4 | 1.17 | 362 | 1.6 | 0.51 | 0.658 | Agricultural and Biological Sciences; Biochemistry, Genetics and Molecular Biology |
| *Communications in Nonlinear Science and Numerical Simulation* | 4 | 1.17 | 1291 | 4.03 | 1.326 | 1.805 | Mathematics |
| *Journal of Crystal Growth* | 4 | 1.17 | 915 | 1.6 | 0.515 | 0.869 | Physics and Astronomy; Chemistry; Materials Science |
| The other 128 journals | 164 | 48.81 | 745 | - | - | - | - |
| Conference papers | 79 | 23.03 | 80 | - | - | - | - |
| **Total** | **343** | **100** | **592** | - | - | - | - |

We have already discussed the issue with *Tumor Biology* and *Diagnostic Pathology*, here above; Table 4 which lists the top 10 journals when most retractions are by Iran-affiliated authors, confirms that these two journals contains papers, which are retracted by many authors from Iranian institutions. Among the top ten journals from Table 4, *Progress in Organic Coatings* had the fastest retraction times (the mean time between publication and retraction being 87 days). This is an amazing result, given the overall mean of almost 600 days. Notice that for *Energy and Buildings*, it takes almost *three thousand days* from publication to retraction day; this journal had the highest CiteScore (5.36) and SJR (1.934) among the top 10 journals with most retractions. (The highest SNIP belonged to *Materials Science and Engineering A*.)

Table 5. Top 10 institutions with most retractions, ordered decreasingly by their share (calculated as the ratio of the number of retracted papers affiliated to the mentioned institution with respect to all retracted papers, i.e., 343). Data source: Scopus.

| Affiliations | No. of documents (TD) | No. of retracted papers (RP) | RP/TD (%) | RP Share | Main reasons for retraction | | | | |
|---|---|---|---|---|---|---|---|---|---|
| | | | | | Fake peer review (%) | Plagiarism (%) | Unspecified (%) | Duplicate publication (%) | Authorship problems (%) |
| **Islamic Azad University** | 78,789 | 182 | 0.23 | 53.1 | 36.8 | 24.1 | 29.1 | 14.8 | 9.3 |
| **University of Tehran** | 56,421 | 49 | 0.09 | 14.3 | 51 | 46.9 | 4 | 12.2 | 34.6 |
| **Baqiyatallah University of Medical Sciences** | 6,080 | 29 | 0.48 | 8.5 | 96.5 | 31 | 3.4 | 6.8 | 31 |
| **AJA University of Medical Sciences** | 1,230 | 25 | 2.03 | 7.3 | 88 | 24 | 4 | 8 | 28 |
| **Tehran University of Medical Sciences** | 49,346 | 22 | 0.04 | 6.4 | 45.4 | 36.6 | 4.5 | 22.7 | 31 |
| **Ilam University** | 1,749 | 15 | 0.86 | 4.4 | 93.3 | 60 | 0 | 6.6 | 60 |
| **Kurdistan University of Medical Sciences** | 2,072 | 15 | 0.72 | 4.4 | 86.6 | 40 | 0 | 6.6 | 33.3 |
| **Urmia University** | 7,297 | 13 | 0.18 | 3.8 | 100 | 53.8 | 0 | 0 | 38.4 |
| **Shahid Beheshti University of Medical Sciences** | 25,808 | 13 | 0.05 | 3.8 | 61 | 30.7 | 0 | 23 | 15.3 |
| **Ferdowsi University of Mashhad** | 17,809 | 12 | 0.07 | 3.5 | 16.6 | 58.3 | 8.3 | 8.3 | 16.6 |

In Table 5, we list the top 10 institutions with most retractions, in a decreasing order as measured by their share, calculated as the ratio between the number of retracted papers by authors affiliated to the mentioned institution and the total number of retracted papers, i.e., 343. Five of the top 10 institutions with most retractions are medical universities (see also Table 3), a result that does not come as a surprise, given that Ghorbi (2019) already pointed out that about half of the retracted Iranian papers dealt with Life Sciences & Biomedicine. Over half (182) of the hereby examined retracted papers were published by authors affiliated with the Islamic Azad University (Table 5). This university has many branches in Iranian cities, and considered together they create a large university; hence so many retracted papers affiliated to it. Among the top 10 listed institutions, however, the share of retracted articles in all articles published by this university is not truly large (0.23%). Other institutions with many retractions include the University of Tehran (49) and Baqiyatallah University of Medical Sciences (29).

In terms of the share of retracted papers to the total number of documents from the institution, AJA University of Medical Sciences stands out, with over 2% of all papers having been retracted—a very high number, even compared to the other top ten institutions (Table 5). The next institutions in this regard are Ilam University (0.86%) and Kurdistan University of Medical Sciences (0.72%). The main retraction reasons of papers from these universities were fake peer review, plagiarism, and authorship problems. Almost all retracted papers from Urmia University, Baqiyatallah University of Medical Sciences, and Ilam University were retracted due to fake peer review; papers from Ilam University, Ferdowsi University of Mashhad, and Urmia University were often retracted due to plagiarism.

Next, consider Figure 6 on which we visualize the collaboration network of the authors of the retracted Iranian papers. Overall, researchers from 27 countries coauthored the retracted Iranian publications; notice that for readability, the graph includes only those countries that shared at least three retracted publications with Iranian authors; without this limitation, the network would include 27 countries. The size of a node size indicates the corresponding collaboration's frequency, while node colors represent continents, yellow representing Asian countries, blue representing European countries, red representing North American countries, and green representing South American countries. Most of the collaborations were with colleagues from Asian and European countries; Malaysia (22 shared publications with Iran) being the greatest foreign contributor to the retracted Iranian papers.

Hayati and Didegah (2010) stated that "Iranian researchers have had scientific collaboration with 115 countries, […] between 1998 and 2007. […]. Iran's main partners were the USA, Canada, and UK." Note that although Iranian researchers cooperate a lot with their colleagues from the USA, Canada, and the UK, these three countries did not appear as key players in a collaboration leading to retraction.

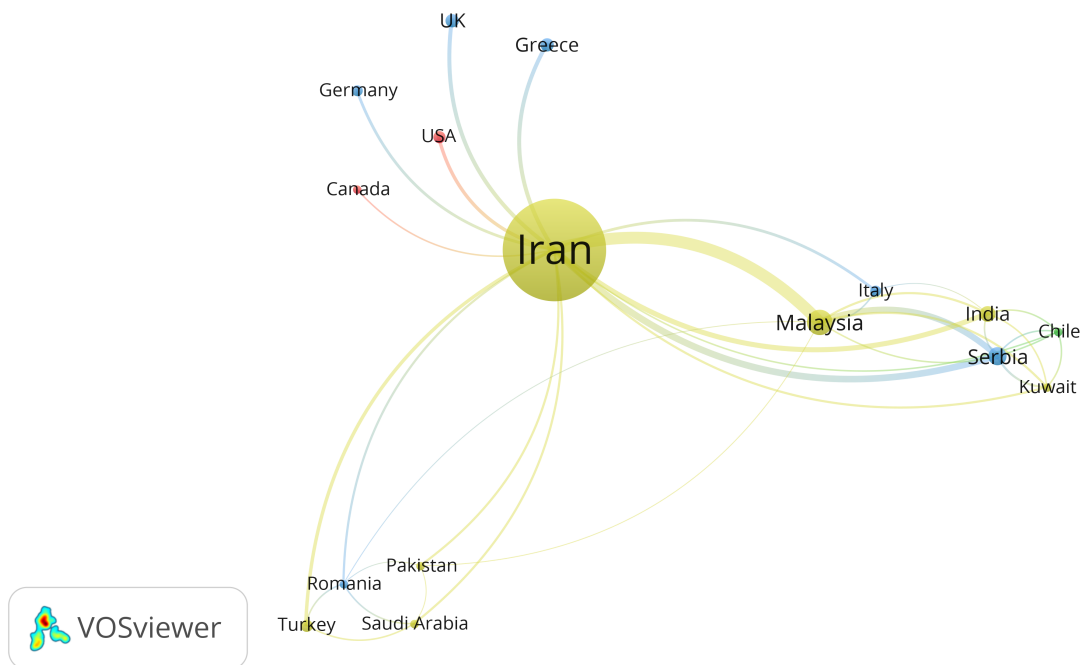

Fig 6. International collaboration network, showing countries that were affiliated in at least three retracted Iranian papers. Without this limitation, the network would include 27 countries. Data source: Scopus.

**Discussion and Conclusion**

In the twenty-first century, science has been developing very quickly, an indication of the overall development of society and humanity in general. Unfortunately, this does not come without cost: researchers are expected to publish much more than two or three decades ago. Working under such an academic pressure, some researchers decide to choose an easy path to publication, one that is based on unethical behavior. Thus, an increasing number of studies have been recognized as publishing false or unethically obtained data. Plagiarism is still a common problem.

These are among the reasons behind the rapidly increasing number of retracted works over the past 20 years (Van Noorden, 2011; Steen et al., 2013). On one hand, this increase in retractions can be considered as good news: more and more unethical papers are being caught; but on the other hand, it is bad news: more and more unethical papers are being published. It is no wonder, then, that many researchers have studied retractions; given the importance of the topic—for science and its development, the scientific community, and society in general—we should pay even more attention to this problem (Van Noorden, 2011; Wager, 2015).

The issue of retractions is difficult to study for various reasons, one of them being the vague meaning of the phenomenon. Take plagiarism; surely, it is wrong, and no one will disagree. Now take retraction; retraction is a tool against wrong behavior, so it is actually a *positive* aspect of the research field features, - although most people will likely have negative thoughts about the phenomenon, as it is closely related to plagiarism and unethical behavior. From a general perspective, an increasing—and already large—number of retractions indicates that there is something significantly wrong with the dissemination of scientific research. In many cases, there are strong reasons to believe that the authors of these articles behaved in an unethical way, and so the articles should not have been published at all. Thus, unfortunately, the retraction phenomenon poses a challenging topic to study, since the number of retractions is in fact a reflection of three confounded factors: *committed unethical behavior*, *caught unethical behavior*, and *false-positive unethical behavior*. Not caught, unethical behavior will not lead to a retraction. Sometimes there will be false positives, that is, retracted articles that should not have been retracted. Of course, there are still published articles resulting directly from unethical behavior.

We want to stress how difficult it is to analyze the retraction phenomenon. A linear approach, reflected by an overly simplified method in which the number of retractions is considered to be a direct measure of unethical behavior, can lead to unfair or ambiguous conclusions. The number of retractions, thus, must be considered within a wider context, and in a more indirect way than the linear approach. That is why we have tried to include as much context and focus as possible in our analyses.

We stress that several articles might have gone unnoticed had they not been retracted, but others have attracted extraordinary attention due to their contents *before* their retraction, something that might have led to additional problems, including societal ones. For example, Wakefield et al. (1998) claimed that combined vaccines made from measles, mumps, and rubella vaccine would cause autism in children. All of that later was proved to be untrue; *The Lancet* retracted this article—and yet, many people worldwide still believe that such vaccinations can cause autism in children; consequently, the vaccination rate of children in the UK and the USA decreased. In another case, two articles on the subject of human stem cells published in *Science* were retracted because the authors fabricated the data (Hwang et al., 2004, 2005). A well-known case about a change in attitudes toward same-sex marriage, also published in *Science*, was retracted because one author failed to provide raw research data (LaCour and Green, 2014). These examples prove that retraction is not efficient: although retracted papers are not considered valid research, their results can outlive their publication, sometimes with long-term effects. Thus, although retraction remains an important tool against unethical behavior, it should be considered as a last resort measure, not the main one.

From the above results, the following picture emerges. Science in the Islamic Republic of Iran suffers from unethical behavior, but this phenomenon, recognized as serious several years ago, seems to be on the decline in recent years. We are aware that analyzing retractions does not mean analyzing all unethical behavior in science—but only the set of behaviors that have been discovered. Moreover, one retracted paper can lead to serious social consequences, such as the reduction in vaccination rates, as discussed above; other retractions can be of lesser significance, and can even occur due to a request from the authors of the retracted paper for unintentional errors.

Thus, an important aspect of retraction is *why* a paper was retracted. However detailed the categorization of retractions is, it will unlikely offer sufficient granularity to easily aggregate these retractions: some causes are indeed "comparable", but others may deserve to be treated in a different or unique way. Imagine that some authors do not provide the data from their article *because no data existed*. Thereafter, imagine other authors may not have been able to provide the data from their article *because of an accident with a computer on which the data were stored*. Are the two situations similar? They definitely are different—which does not mean that the journal could not or should not retract both articles, in particular because no one besides the authors actually knows the truth. It is difficult to imagine the former authors to say, "Sorry, but we falsified the data." They would rather choose the latter explanation, just like the latter authors. As we see, studying retractions and other types of unethical behavior in any discipline is complex. When using general data, like those we used here, we have to keep this lack of unquestionable truth in mind when interpreting data on retractions.

Here, we have studied Iranian retractions because, in the past, the Islamic Republic of Iran has been pointed out as one of the greatest contributors to the retraction fields. Our study confirms that Iran indeed plays a role, but also that this role seems less drastic than it was several years ago. This may be due to the actions, including degradation in academic rankings and even expulsion from research centers, that were taken in the country after reports of serious unethical behavior by Iranian scientific authors, as described by Dakhesh and Hamidi (2020).

The study covered works by authors affiliated with at least one Iranian institution, published over 2001–2019. Over this period, the number of retracted articles by Iranian authors increased during 2001–2019. This trend was interrupted by two peaks, in 2010 and 2016. In fact, those two years contributed to over 50% of all the retractions from the period studied. Thus, the total number of documents was steadily growing, but the number of retracted papers rather grew erratically. If confirmed in the following years, this trend may mean that the actions which the Islamic Republic of Iran took against unethical scientists have been fruitful.

The mean time from publication to retraction was almost 600 days (about 20 months); the median was 338 days. It was much shorter for conference papers than for journal papers. Sometimes retractions happened very quickly (less than 100 days from publication), but sometimes they occurred after several years. For example, it took over eight years to retract (five) publications from *Energy and Buildings*. Surprising at first glance, this observation is ''understandable'' because the papers are duplicate publications. This example of delayed resolution indicates that unethical behavior can be discovered long after the act.

Our results show that none of the retractions from conference papers provided a reason for the retraction. Unfortunately, our data analysis does not give any indication as to why these reasons were not provided—but given the scale of the retractions, it seems that one should be worrying why retraction reasons from conference proceedings were not specified. This is something that should be stressed—and changed. Journals, on the other hand, often publish reasons behind retractions, often publishing an erratum to the retracted paper.

In our sample, 264 retracted papers were published in 138 different journals; the remaining 79 papers were published in conference proceedings. However, about 30% of the retracted Iranian articles were published in 10 journals, and 14% of the retracted articles were published in *Tumor Biology* and *Diagnostic Pathology*, both being medical journals. This shows that large-scale retractions happen. This also shows that the scholarly publishing community is often brave enough to take serious actions against unethical behavior.

In an ideal world, no article would need retraction because no article would deserve it. As we do not live in such a world, the scientific community is not perfect. So, As scientists, we need to react when an article presents such problems. We must remember that although editors are among the people responsible for the quality of the papers published in a particular journal, the current peer review system is strongly based upon scientists who serve as reviewers. The quality of reviewers is a known challenging criterion of the peer review process (Ausloos et al., 2016). So if an article has actual ''errors'', it is often due to how the system works—sometimes an editor will fail, sometimes a reviewer. The system is far from perfect, as the growing retraction phenomenon confirms. It takes courage to retract multiple articles from one journal, and the size and reputation of the journal does not matter. Often it may be much easier to ignore the problem. The editors and publishers who decide to act, particularly when they retract many papers simultaneously, deserve the highest praise.

In our results, among the reasons for the retracted articles, those related to peer review seem to play the most important role, with fake peer review being the most frequent retraction reason, followed by plagiarism. This part of our results is not consistent with the research of Ribeiro and Vasconcelos (2018), Samp et al. (2012), Fanelli (2009), and Steen (2011), who identified "other reasons" as the most common reason for retraction—although it should be noted that this term, "other reasons", does not explain the actual reasons. However, we are not the only ones to indicate plagiarism as an important retraction reason; see Chauvin et al. (2019), Elango et al. (2019), and Rubbo et al. (2019).

The analysis of institutions resulted in the Islamic Azad University as that which contributed most to Iranian retractions. Ghorbi and Fahimifar (2020) reported the same observation. However, since researchers affiliated to this (huge) university publish much, this result is somewhat expected. Such a raw-data analysis can be misleading; thus, we normalized the data by calculating the percentage of retracted papers to all papers published from the institution. After this normalization, the Islamic Azad University, with only 0.25% of all articles being retracted, was no longer the most often retracted Iranian institution; it is replaced by AJA University of Medical Sciences. The raw number of 25 retracted articles from the latter university perhaps does not draw attention, but the share of these 25 articles in all the articles from this university does— over 2%.

Out of the 343 retracted papers, only 64 resulted from cooperation with colleagues affiliated with institutions from other countries. (Note that this does not mean that they were not Iranian, since we did not study their nationalities, but only affiliations.) Among those 64 papers, most were affiliated with Asian and European countries, with Malaysia having the largest share. This goes against the presumably intuitive expectation as well as the findings by Elango et al. (2019) and Steen (2011), in which studies in collaboration with colleagues from the USA were predominant.

All in all, the results suggest that fake peer review plays a pivotal role in the retractions of Iranian articles, affiliated not only with lesser known but also with the most prestigious Iranian scientific institutions, and published in quite reputable international journals. This being the case, the peer review system emerges as the most important measure of the fight against unethical behavior, but also as the most susceptible measure to failure. The scholarly publishing community has been using this system for decades, but what worked 30 years ago does not necessarily work now, when the scientific community is working under completely different levels of competition, expectations, and the resulting pressure to publish. So, the increasing trend in retractions does suggest that something must be changed. New peer review systems, which we can see applied in various (although still not numerous) journals, will perhaps offer a solution; but still, we are not there.

In terms of retractions of Iranian publications, we are seeing a gradual improvement. Although this country has strongly contributed to overall global retractions, the most recent years have shown an improvement. Of course, we still face an increasing number of retractions of Iranian papers, but this trend is not as steep, and it seems that Iran does not stand out anymore. Two previous years (2010 and 2016) were significantly bad in this regard, with so the most retracted papers. We have to hope that in the following years the situation will stabilize and even improve.

**Limitations**

As with any study, ours has some limitations. They include the following phenomena and issues.

*Lacking of justification for retractions*

Some retractions did not provide any justification. None of the retracted conference papers provided a reason. This lack of transparency in retraction reasons hinders the analysis of this important aspect of the phenomenon. Unfortunately, this limitation cannot be overcome in such studies as ours. To study this aspect of retractions, a more in-depth qualitative study is needed. It could, for example, include interviews with the editors who retracted some articles (like a set of conference proceedings) but did not provide the reasons behind this decision.

*Lack of information on those responsible for unethical behavior*

It is not possible to determine with certainty which among the authors of a retracted paper were involved in misconduct resulting in the article retraction. This issue also calls for a dedicated qualitative study, but this time among the authors of retracted papers. This is a very interesting topic, because it is possible that some of the authors of retracted papers were not aware of the unethical behavior of others. Such knowledge could help gain additional knowledge about unethical behavior in science, which could help in future efforts at prevention. Also, it could help scientists learn how to avoid being victims of unethical behaviors by their co-authors.

That said, this topic is extremely difficult to study, and the responses one would get from participants would be extremely difficult to verify. Hence, such a study could employ methods used in survey sampling to study sensitive data.

*Retraction weight*

As discussed above, two retractions with the same reason do not necessarily bear the same ethical weight. A quantitative study like ours equalizes all retractions that were assigned a particular reason, an approach that can miss important differences between retractions. One author might have done something wrong intentionally—which makes it unethical—while another might have made an accidental error—which means the behavior was *not* unethical. Again, studying this topic calls for a qualitative study of what has happened in particular retractions, which could require investigating each of the situations that led to the retractions studied.

*Limitations of data sources*

An obvious but unavoidable limitation are the data sources themselves. Search engines can provide inconsistent results (Aspura et al., 2018). We have examined data based on Scopus and Retraction Watch. These two different data sources do not offer directly comparable data; on the other hand, studying two or more data sources can help to draw a more representative picture of the phenomenon.

**Future research direction**

A study of retracted papers is complex and draws a wide picture of the phenomenon, in particular of Iranian retractions, but by no means does it end the research objectives. In addition to the various ideas presented above in the limitations, further research can be related to citations of retracted articles. Another line of research can compare retracted and non-retracted papers in terms of citations, MCC, international collaboration, etc.

An interesting topic is related to research and ethical guidelines in scientific institutions; in particular, those related to scientific misconduct, the obligations authors must obey when conducting research, and how offenders would be and have been treated based on these guidelines. Such a study, when correlated with retractions (and other types of punishment after misconduct) from these organizations, could offer insights into whether such guidelines are effective. Of course, it would be also interesting to study whether scientific institutions that have such rules obey them.

Studies based on quantitative data offer a lot of information, but this information is limited to what is present in the data. Particularly interesting and valuable research would be to determine the understanding and intentions of authors whose papers were retracted. What did you really do, and how serious was it? Why did you do it? Do you consider yourself a scientific criminal? Would you do it again? How did it affect your career? Do you regret this? Such a qualitative study could provide a world of new knowledge on unethical behavior, retractions, and scientists.

**Appendix**

As discussed in the Results section, in 2016, *Tumor Biology* and *Diagnostic Pathology* retracted many Iranian papers. The figure below shows the number of Iranian papers these two journals published in 2004–2020.

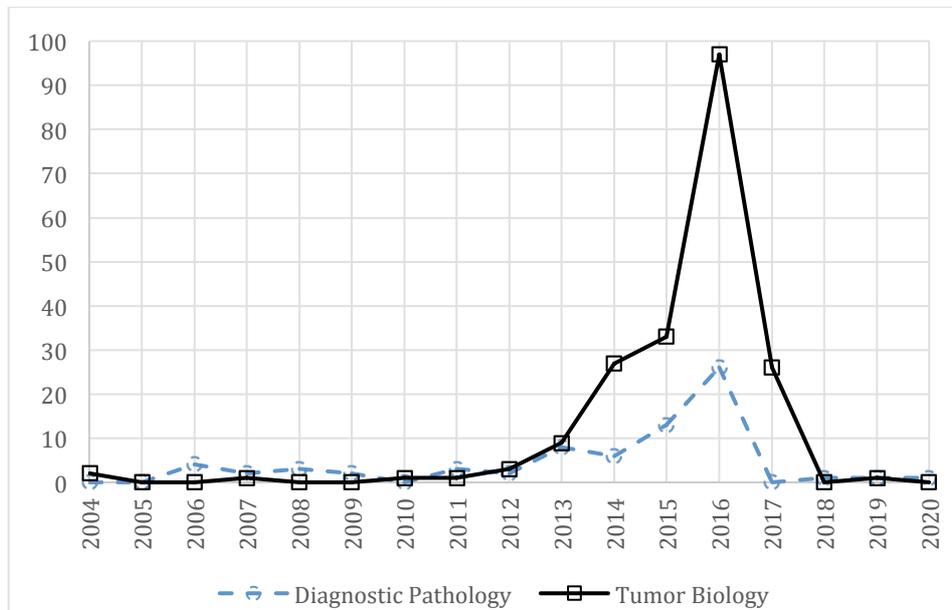

Fig A1. The number of Iranian publications published by *Diagnostic Pathology* and *Tumor Biology*. Data source: Scopus

# References

Ajiferuke, I., Burell, Q., & Tague, J. (1988). Collaborative coefficient: A single measure of the degree of collaboration in research. *Scientometrics, 14*(5-6), 421-433, doi: 10.1007/bf02017100.

Aspura, M. Y. I., Noorhidawati, A., & Abrizah, A. (2018). An analysis of Malaysian retracted papers: Misconduct or mistakes? *Scientometrics, 115*(3), 1315-1328, doi: 10.1007/s11192-018-2720-z .

Ausloos, M., Nedic, O., & Dekanski, A. (2016). Day of the week effect in paper submission/acceptance/rejection to/in/by peer review journals. *Physica A: Statistical Mechanics and its Applications, 456*, 197-203, doi: 10.1016/j.physa.2016.03.032.

Bornmann, L., & Mungra, P. (2011). Improving peer review in scholarly journals. *European Science Editing, 37*(2), 41-43.

Bornemann-Cimenti, H., Szilagyi, I. S., & Sandner-Kiesling, A. (2016). Perpetuation of Retracted Publications Using the Example of the Scott S. Reuben Case: Incidences, Reasons and Possible Improvements. *Science and Engineering Ethics, 22(4)*, 1063-1072.

Budd, J. M., Sievert, M. E., & Schultz, T. R. (1998). Phenomena of retraction - Reasons for retraction and citations to the publications. *Jama-Journal of the American Medical Association, 280*(3), 296-297, doi: 10.1001/jama.280.3.296.

Callaway E. (2016). Publisher pulls 58 articles by Iranian scientists over authorship manipulation. Nature. (Accessed 01/15/2019.) Available at: https://www.nature.com/news/publisher-pulls-58-articles-by-iranian-scientists-over-authorship-manipulation-1.20916.

Campos-Varela, I., & Ruano-Raviña, A. (2019). Misconduct as the main cause for retraction. A descriptive study of retracted publications and their authors. *Gaceta Sanitaria, 33*, 356-360, doi: 10.1016/j.gaceta.2018.01.009.

Chauvin, A., De Villelongue, C., Pateron, D., & Yordanov, Y. (2019). A systematic review of retracted publications in emergency medicine. *European Journal of Emergency Medicine, 26*(1), 19-23, doi: 10.1097/mej.0000000000000491.

Dakhesh, S., & Hamidi, A. (2020). Scientific misconduct and Iranian scientists. *Gaceta Sanitaria, 33*, 598-598, doi: 10.1016/j.gaceta.2019.02.003.

Dal-Ré, R., & Ayuso, C. (2019). Reasons for and time to retraction of genetics articles published between 1970 and 2018. *Journal of Medical Genetics*, 56, 734-740, doi: 10.1136/jmedgenet-2019-106137.

De Solla Price, D. J., & Beaver, D. (1966). Collaboration in an invisible college. *American Psychologist, 21*(11), 1011–1018. doi.org/10.1037/h0024051

Elango, B., Kozak, M., & Rajendran, P. (2019). Analysis of retractions in Indian science. *Scientometrics, 119*(2), 1081-1094, doi: 10.1007/s11192-019-03079-y.

Fanelli, D. (2009). How Many Scientists Fabricate and Falsify Research? A Systematic Review and Meta-Analysis of Survey Data. *PLoS ONE, 4*(5), e5738, doi: 10.1371/journal.pone.0005738.

Fang, F. C., Steen, R. G., & Casadevall, A. (2012). Misconduct accounts for the majority of retracted scientific publications. *Proceedings of the National Academy of Sciences of the United States of America, 109*(42), 17028-17033, doi: 10.1073/pnas.1212247109.

Ghorbi, A. (2019). Examine the aspects of research retraction and provide guidelines for identifying unreliable publications. (Master's thesis), University of Tehran, Tehran, Iran. https://thesis2.ut.ac.ir/thesis/UTCatalog/UTThesis/Forms/ThesisBrief.aspx?thesisID=d26eb04f-7aa6-4217-833b-efb057f16db8

Ghorbi, A., Fahimifar, S. (2020). Aspects and Collaboration Patterns of Retracted Papers as Evidence of Research Misconduct in Iran and Foreign countries. *Journal of Scientometrics*, 6(11), 149-172. doi: 10.22070/rsci.2019.4392.1287.

Grieneisen, M. L., & Zhang, M. (2012). A Comprehensive Survey of Retracted Articles from the Scholarly Literature. *PLoS ONE, 7*(10), e44118, doi: 10.1371/journal.pone.0044118.

Hayati, Z., & Didegah, F. (2010). International scientific collaboration among Iranian researchers during 1998-2007. *Library Hi Tech*, 28(3), 433-446, doi: 10.1108/07378831011066675.

Kozak, M. (2008). Correlation coefficient and the fallacy of statistical hypothesis testing. *Current Science*, 95(9), 1121-1122.

Lei, L., & Zhang, Y. (2018). Lack of Improvement in Scientific Integrity: An Analysis of WoS Retractions by Chinese Researchers (1997–2016). *Journal of Science and Engineering Ethics, 24*(5), 1409-1420, doi:

Ribeiro, M. D., & Vasconcelos, S. M. R. (2018). Retractions covered by Retraction Watch in the 2013–2015 period: prevalence for the most productive countries. *Scientometrics, 29*(8), 719-734, doi: 10.1007/s11192-017-2621-6.

Rubbo, P., Helmann, C. L., dos Santos, C. B., & Pilatti, L. A. (2019). Retractions in the Engineering Field: A Study on the Web of Science Database. *Ethics and Behavior, 29*(2), 141-155, doi: 10.1080/10508422.2017.1390667.

Samp, J. C., Schumock, G. T., & Pickard, A. S. (2012). Retracted Publications in the Drug Literature. *Pharmacotherapy, 32*(7), 586-595, doi: 10.1002/j.1875-9114.2012.01100.x.

Savanur, K., & Srikanth, R. (2010). Modified collaborative coefficient: A new measure for quantifying the degree of research collaboration. *Scientometrics*, 84(2), 365-371, doi: 10.1007/s11192-009-0100-4.

Stavale, R., Ferreira, G. I., Galvão, J. A. M., Zicker, F., Novaes, M. R. C. G., de Oliveira, C. M., & Guilhem, D. (2019). Research misconduct in health and life sciences research: A systematic review of retracted literature from Brazilian institutions. *PLoS one*, *14*(4), e0214272, doi: 10.1371/journal.pone.0214272.

Steen, R. G. (2011). Retractions in the scientific literature: is the incidence of research fraud increasing? *Journal of Medical Ethics, 37*, 249-253, doi: 10.1136/jme.2010.040923.

Steen, R. G., Casadevall, A., & Fang, F. C. (2013). Why Has the Number of Scientific Retractions Increased? *PLoS ONE, 8*, e68397, doi: 10.1371/journal.pone.0068397.

Steen, R. G. (2012). Retractions in the medical literature: how can patients be protected from risk?. *Journal of medical ethics*, *38*(4), 228-232.

Stewart, W. W., & Feder, N. (1987). The integrity of the scientific literature. *Nature*, *325*, 207-214.

Tang, L., Hu, G., Sui, Y., Yang, Y., & Cao, C. (2020). Retraction: The "Other Face" of Research Collaboration?. *Science and Engineering Ethics*, 26, 1681-1708, doi: 10.1007/s11948-020-00209-1.

Van Noorden, R. (2011). Science publishing: The trouble with retractions. *Nature, 478*(7367), 26-28, doi: 10.1038/478026a.

Vuong, Q. H. (2020). The limitations of retraction notices and the heroic acts of authors who correct the scholarly record: An analysis of retractions of papers published from 1975 to 2019. *Learned Publishing*, *33*(2), 119-130, doi: 10.1002/leap.1282.

Wager, E. (2015). Why are retractions so difficult? *Science Editing, 2*(1), 32-34, doi: 10.6087/kcse.34.

Wager, E., Barbour, V., Yentis, S., & Kleinert, S., (2009). Retractions: Guidance from the Committee on Publication Ethics (COPE). *Croatian Medical Journal, 50*(6), 532–535, doi: 10.3325/cmj.2009.50.532.

Wager, E., & Williams, P. (2011). Why and how do journals retract articles? An analysis of Medline retractions 1988-2008. *Journal of Medical Ethics, 37*(9), 567-570, doi: 10.1136/jme.2010.040964.

Zhang, C., Ding, K., & Liu, Z. (2019). Informetric Analysis on the International Retracted Publication Based on the Web of Science Database. *Advances in Social Science, Education and Humanities Research, 376* (), 576-585, doi: 10.2991/sschd-19.2019.100.

**References to retracted papers**

Hwang, W. S., Roh, S. I., Lee, B. C., Kang, S. K., Kwon, D. K., Kim, S., et al. (2005). Patient-specific embryonic stem cells derived from human SCNT blastocysts. *Science, 308*(5729), 1777-1783, doi: 10.1126/science.1112286.

Hwang, W. S., Ryu, Y. J., Park, J. H., Park, E. S., Lee, E. G., Koo, J. M., ... & Moon, S. Y. (2004). Evidence of a pluripotent human embryonic stem cell line derived from a cloned blastocyst. *Science, 303*(5664), 1669-1674.

LaCour, M. J., & Green, D. P. (2014). When contact changes minds: An experiment on transmission of support for gay equality. *Science, 346*(6215), 1366-1369, doi: 10.1126/science.1256151

Wakefield, A. J., Murch, S. H., Anthony, A., Linnell, J., Casson, D. M., Malik, M., et al. (1998). Ileal-lymphoid-nodular hyperplasia, non-specific colitis, and pervasive developmental disorder in children. *Lancet, 351*(9103), 637-641, doi: 10.1016/s0140-6736(97)11096-0.